\documentstyle[12pt]{article}

\def\doit#1#2{\ifcase#1\or#2\fi}

\skewchar\fivmi='177
\skewchar\sixmi='177
\skewchar\sevmi='177
\skewchar\egtmi='177
\skewchar\ninmi='177
\skewchar\tenmi='177
\skewchar\elvmi='177
\skewchar\twlmi='177
\skewchar\frtnmi='177
\skewchar\svtnmi='177
\skewchar\twtymi='177
\def\@magscale#1{ scaled \magstep #1}

\def\framingfonts#1{
\doit{#1}{\font\twfvmi  = ammi10   \@magscale5 
\skewchar\twfvmi='177
\skewchar\fivsy='60
\skewchar\sixsy='60
\skewchar\sevsy='60
\skewchar\egtsy='60
\skewchar\ninsy='60
\skewchar\tensy='60
\skewchar\elvsy='60
\skewchar\twlsy='60
\skewchar\frtnsy='60
\skewchar\svtnsy='60
\skewchar\twtysy='60
\font\twfvsy  = amsy10   \@magscale5 
\skewchar\twfvsy='60
\font\go=font018			
\font\sc=font005			
\def\Go#1{{\hbox{\go #1}}}	
\def\Sc#1{{\hbox{\sc #1}}}	
\def\Sf#1{{\hbox{\sf #1}}}	
\font\oo=circlew10	      
\font\ooo=circle10			
\font\ro=manfnt				
\def\kcl{{\hbox{\ro 6}}}		
\def\kcr{{\hbox{\ro 7}}}		
\def\ktl{{\hbox{\ro \char'134}}}	
\def\ktr{{\hbox{\ro \char'135}}}	
\def\kbl{{\hbox{\ro \char'136}}}	
\def\kbr{{\hbox{\ro \char'137}}}	
}}

\catcode`@=11
\def\un#1{\relax\ifmmode\@@underline#1\else
	$\@@underline{\hbox{#1}}$\relax\fi}
\catcode`@=12

\let\du=\d			
\let\um=\H			

\def\d{\delta}
\def\e{\epsilon}

\def\m{\mu}

\def\plpl{{\raise-2pt\hbox{$\raise3pt\hbox{$_+$}\hskip-7.0pt\raise0.0pt
\hbox{$^+$}\hskip 0.01pt$}}}
\def\mimi{{\raise-2pt\hbox{$\raise3pt\hbox{$_-$}\hskip-7.0pt\raise0.0pt
\hbox{$^-$}\hskip 0.01pt$}}}

\def\bo{{\raise.15ex\hbox{\large$\Box$}}}		
\def\TH{{\raise.2ex\hbox{$\displaystyle \bigodot$}\mskip-4.7mu \llap H \;}}
\def\face{{\raise.2ex\hbox{$\displaystyle \bigodot$}\mskip-2.2mu \llap {$\ddot
	\smile$}}}					

\def\sp#1{{}^{#1}}				
\def\Tilde#1{{\widetilde{#1}}\hskip 0.03in}			
\def\Bar#1{\overline{#1}}			
\def\leftrightarrowfill{$\mathsurround=0pt \mathord\leftarrow \mkern-6mu
	\cleaders\hbox{$\mkern-2mu \mathord- \mkern-2mu$}\hfill
	\mkern-6mu \mathord\rightarrow$}
\def\dvec#1{\vbox{\ialign{##\crcr
	\leftrightarrowfill\crcr\noalign{\kern-1pt\nointerlineskip}
	$\hfil\displaystyle{#1}\hfil$\crcr}}}		
\def\dt#1{{\buildrel {\hbox{\LARGE .}} \over {#1}}}	

\def\frac#1#2{{\textstyle{#1\over\vphantom2\smash{\raise.20ex
	\hbox{$\scriptstyle{#2}$}}}}}			
\def\sfrac#1#2{{\vphantom1\smash{\lower.5ex\hbox{\small$#1$}}\over
	\vphantom1\smash{\raise.4ex\hbox{\small$#2$}}}}	
\def\bfrac#1#2{{\vphantom1\smash{\lower.5ex\hbox{$#1$}}\over
	\vphantom1\smash{\raise.3ex\hbox{$#2$}}}}	
\def\afrac#1#2{{\vphantom1\smash{\lower.5ex\hbox{$#1$}}\over#2}}    

\newskip\humongous \humongous=0pt plus 1000pt minus 1000pt
\def\caja{\mathsurround=0pt}
\def\eqalign#1{\,\vcenter{\openup2\jot \caja
	\ialign{\strut \hfil$\displaystyle{##}$&$
	\displaystyle{{}##}$\hfil\crcr#1\crcr}}\,}
\newif\ifdtup
\def\panorama{\global\dtuptrue \openup2\jot \caja
	\everycr{\noalign{\ifdtup \global\dtupfalse
	\vskip-\lineskiplimit \vskip\normallineskiplimit
	\else \penalty\interdisplaylinepenalty \fi}}}
\def\li#1{\panorama \tabskip=\humongous				
	\halign to\displaywidth{\hfil$\displaystyle{##}$
	\tabskip=0pt&$\displaystyle{{}##}$\hfil
	\tabskip=\humongous&\llap{$##$}\tabskip=0pt
	\crcr#1\crcr}}

\def\ref#1{$\sp{#1)}$}

\parskip=\medskipamount			
\thicklines			    

\def\oldheadpic{				
	\setlength{\unitlength}{.4mm}
	\thinlines
	\par
	\begin{picture}(349,16)
	\put(325,16){\line(1,0){4}}
	\put(330,16){\line(1,0){4}}
	\put(340,16){\line(1,0){4}}
	\put(335,0){\line(1,0){4}}
	\put(340,0){\line(1,0){4}}
	\put(345,0){\line(1,0){4}}
	\put(329,0){\line(0,1){16}}
	\put(330,0){\line(0,1){16}}
	\put(339,0){\line(0,1){16}}
	\put(340,0){\line(0,1){16}}
	\put(344,0){\line(0,1){16}}
	\put(345,0){\line(0,1){16}}
	\put(329,16){\oval(8,32)[bl]}
	\put(330,16){\oval(8,32)[br]}
	\put(339,0){\oval(8,32)[tl]}
	\put(345,0){\oval(8,32)[tr]}
	\end{picture}
	\par
	\thicklines
	\vskip.2in}
\def\oldtitle#1#2#3#4{\oldheadpic\begin{center}\vglue.5in{\large\bf #1}\\[.6in]
	{#2}\\[.1in] {\it Department of Physics and Astronomy}\\
	{\it University of Maryland, College Park, MD 20742}\\[.6in]
	Physics Publication \#{#3}\\ {#4}\\[1.5in] {\bf Abstract}\\[.1in]
	\end{center} \begin{quotation}}			
\def\oldTitle#1#2#3#4#5#6#7{\oldheadpic\begin{center} \vglue .4in
	{\large\bf #1}\\[.4in]
	{#2}\\[.1in] {\it Department of Physics and Astronomy}\\
	{\it University of Maryland, College Park, MD 20742}\\[.1in]
	{#3}\\[.1in] {\it {#4}}\\ {\it {#5}}\\[.4in]
	Physics Publication \#{#6}\\ {#7}\\[.5in] {\bf Abstract}\\[.1in]
	\end{center} \begin{quotation}}			
\def\border{						
	\setlength{\unitlength}{1mm}
	\newcount\xco
	\newcount\yco
	\xco=-24
	\yco=12
	\begin{picture}(140,0)
	\put(\xco,\yco){$\ktl$}
	\advance\yco by-1
	{\loop
	\put(\xco,\yco){$\kcl$}
	\advance\yco by-2
	\ifnum\yco>-240
	\repeat
	\put(\xco,\yco){$\kbl$}}
	\xco=158
	\yco=12
	\put(\xco,\yco){$\ktr$}
	\advance\yco by-1
	{\loop
	\put(\xco,\yco){$\kcr$}
	\advance\yco by-2
	\ifnum\yco>-240
	\repeat
	\put(\xco,\yco){$\kbr$}}
        \put(-20,11){\tiny University of Maryland Elementary Particle
Physics University of Maryland Elementary Particle Physics University of
Maryland Elementary Particle Physics}
	\put(-20,-241.5){\tiny University of Maryland Elementary
Particle Physics University of Maryland Elementary Particle Physics
University of Maryland Elementary Particle Physics}
	\end{picture}
	\par\vskip-8mm}
\def\bordero{						
	\setlength{\unitlength}{1mm}
	\newcount\xco
	\newcount\yco
	\xco=-24
	\yco=12
	\begin{picture}(140,0)
	\put(\xco,\yco){$\ktl$}
	\advance\yco by-1
	{\loop
	\put(\xco,\yco){$\kcl$}
	\advance\yco by-2
	\ifnum\yco>-240
	\repeat
	\put(\xco,\yco){$\kbl$}}
	\xco=158
	\yco=12
	\put(\xco,\yco){$\ktr$}
	\advance\yco by-1
	{\loop
	\put(\xco,\yco){$\kcr$}
	\advance\yco by-2
	\ifnum\yco>-240
	\repeat
	\put(\xco,\yco){$\kbr$}}
	\put(-20,12){\ooo
bacdefghidfghghdhededbihdgdfdfhhdheidhdhebaaahjhhdahbahgdedgehgfdiehhgdigicba}
	\put(-20,-241.5){\ooo
ababaighefdbfghgeahgdfgafagihdidihiidhiagfedhadbfdecdcdfagdcbhaddhbgfchbgfdacfediacbabab}
	\end{picture}
	\par\vskip-8mm}
\def\headpic{						
	\indent
	\setlength{\unitlength}{.4mm}
	\thinlines
	\par
	\begin{picture}(29,16)
	\put(165,16){\line(1,0){4}}
	\put(170,16){\line(1,0){4}}
	\put(180,16){\line(1,0){4}}
	\put(175,0){\line(1,0){4}}
	\put(180,0){\line(1,0){4}}
	\put(185,0){\line(1,0){4}}
	\put(169,0){\line(0,1){16}}
	\put(170,0){\line(0,1){16}}
	\put(179,0){\line(0,1){16}}
	\put(180,0){\line(0,1){16}}
	\put(184,0){\line(0,1){16}}
	\put(185,0){\line(0,1){16}}
	\put(169,16){\oval(8,32)[bl]}
	\put(170,16){\oval(8,32)[br]}
	\put(179,0){\oval(8,32)[tl]}
	\put(185,0){\oval(8,32)[tr]}
	\end{picture}
	\par\vskip-6.5mm
	\thicklines}
\def\title#1#2#3#4{\border\headpic {\hbox to\hsize{#4 \hfill UMDEPP #3}}\par
	\begin{center} \vglue .5in {\large\bf #1}\\[.6in]
	{#2}\\[.1in] {\it Department of Physics and Astronomy}\\
	{\it University of Maryland, College Park, MD 20742}\\[1.5in]
	{\bf Abstract}\\[.1in] \end{center} \begin{quotation}}	
\def\Title#1#2#3#4#5#6#7{\border\headpic
	{\hbox to\hsize{#7 \hfill UMDEPP #6}}\par
	\begin{center} \vglue .4in {\large\bf #1}\\[.4in]
	{#2}\\[.1in] {\it Department of Physics and Astronomy}\\
	{\it University of Maryland, College Park, MD 20742}\\[.1in]
	{#3}\\[.1in] {\it {#4}}\\ {\it {#5}}\\[.5in] {\bf Abstract}\\[.1in]
	\end{center} \begin{quotation}}			
\def\endtitle{\end{quotation}\newpage}			

\def\sect#1{\bigskip\medskip \goodbreak \noindent{\bf {#1}} \nobreak \medskip}
\def\refs{\sect{References} \footnotesize \frenchspacing \parskip=0pt}
\def\Item{\par\hang\textindent}

\def\doit#1#2{\ifcase#1\or#2\fi}
\def\[{\lfloor{\hskip 0.35pt}\!\!\!\lceil}
\def\]{\rfloor{\hskip 0.35pt}\!\!\!\rceil}

\def\du#1#2{_{#1}{}^{#2}}

\def\hati{{\hat{I}}}
\def\dt{$~D=10$~}

\def\pl#1#2#3{Phys.~Lett.~{\bf {#1}B} (19{#2}) #3}
\def\np#1#2#3{Nucl.~Phys.~{\bf B{#1}} (19{#2}) #3}

\def\prep#1#2#3{Phys.~Rep.~{\bf {#1}C} (19{#2}) #3}

\def\ijmp#1#2#3{Int.~Jour.~Mod.~Phys.~{\bf {#1}} (19{#2}) #3}

\def\ula{{\un a}}
\def\ulb{{\un b}}
\def\ulc{{\un c}}
\def\uld{{\un d}}

\def\fracmm#1#2{{{#1}\over{#2}}}
\def\gg{{\hbox{\sc g}}}
\def\half{{\fracm12}}

\def\frac#1#2{{\textstyle{#1\over\vphantom2\smash{\raise -.20ex
	\hbox{$\scriptstyle{#2}$}}}}}			

\def\fracm#1#2{\hbox{\large{${\frac{{#1}}{{#2}}}$}}}

\def\Dot#1{\buildrel{_{_{\hskip 0.01in}\bullet}}\over{#1}}
\def\dt#1{\Dot{#1}}
\def\uln{{\underline n}}
\def\Tilde#1{{\widetilde{#1}}\hskip 0.015in}

\def\scst{\scriptstyle}
\def\itrema{$\ddot{\scriptstyle 1}$}
\def\Bo{\bo{\hskip 0.03in}}
\def\lrad#1{ \left( A {\buildrel\leftrightarrow\over D}_{#1} B\right) }
\def\derx{\partial_x} \def\dery{\partial_y} \def\dert{\partial_t}
\def\Vec#1{{\overrightarrow{#1}}}
\def\.{.$\,$}

\def\grg#1#2#3{Gen.~Rel.~Grav.~{\bf{#1}} (19{#2}) {#3} }
\def\pla#1#2#3{Phys.~Lett.~{\bf A{#1}} (19{#2}) {#3}}

\def\ula{{\underline a}}
\def\ulb{{\underline b}}
\def\ulc{{\underline c}}
\def\uld{{\underline d}}
\def\ule{{\underline e}}
\def\ulf{{\underline f}}
\def\ulg{{\underline g}}
\def\ulm{{\underline m}}
\def\uln#1{\underline{#1}}
\def\ulp{{\underline p}}
\def\ulq{{\underline q}}
\def\ulr{{\underline r}}

\def\hatm{\hat m}
\def\hatn{\hat n}
\def\hatr{\hat r}
\def\hats{\hat s}
\def\hatt{\hat t}

\def\ul{\underline}
\def\un{\underline}
\def\-{{\hskip 1.5pt}\hbox{-}}

\def\kd#1#2{\d\du{#1}{#2}}
\def\fracmm#1#2{{{#1}\over{#2}}}
\def\footnotew#1{\footnote{\hsize=6.5in {#1}}}

\def\low#1{{\raise -3pt\hbox{${\hskip 1.0pt}\!_{#1}$}}}

\def\ip{{=\!\!\! \mid}}
\def\unb{{\underline {\bar n}}}
\def\upb{{\underline {\bar p}}}
\def\um{{\underline m}}
\def\up{{\underline p}}
\def\Phib{{\Bar \Phi}}
\def\Phit{{\tilde \Phi}}
\def\Phibt{{\tilde {\Bar \Phi}}}
\def\Db{{\Bar D}_{+}}
\def\gg{{\hbox{\sc g}}}
\def\nt{$~N=2$~}

\def\Dot#1{\buildrel{_{_{\hskip 0.01in}\bullet}}\over{#1}}
\def\dt#1{\Dot{#1}}
\def\gg{{\hbox{\sc g}}}
\def\nt{$~N=2$~}
\def\gg{{\hbox{\sc g}}}
\def\nt{$~N=2$~}
\def\tr{{\rm tr}}
\def\Tr{{\rm Tr}}
\def\mpl#1#2#3{Mod.~Phys.~Lett.~{\bf A{#1}} (19{#2}) #3}
\def\hati{{\hat i}} \def\hatj{{\hat j}} \def\hatk{{\hat k}}
\def\hatl{{\hat l}}

\begin{document}

\font\tenmib=cmmib10
\font\sevenmib=cmmib10 at 7pt 
\font\fivemib=cmmib10 at 5pt  
\font\tenbsy=cmbsy10
\font\sevenbsy=cmbsy10 at 7pt 
\font\fivebsy=cmbsy10 at 5pt  
\def\BMfont{\textfont0\tenbf \scriptfont0\sevenbf
                              \scriptscriptfont0\fivebf
            \textfont1\tenmib \scriptfont1\sevenmib
                               \scriptscriptfont1\fivemib
            \textfont2\tenbsy \scriptfont2\sevenbsy
                               \scriptscriptfont2\fivebsy}
\def\rlx{\relax\leavevmode}
\def\BM#1{\rlx\ifmmode\mathchoice
                      {\hbox{$\BMfont#1$}}
                      {\hbox{$\BMfont#1$}}
                      {\hbox{$\scriptstyle\BMfont#1$}}
                      {\hbox{$\scriptscriptstyle\BMfont#1$}}
                 \else{$\BMfont#1$}\fi}

\font\tenmib=cmmib10
\font\sevenmib=cmmib10 at 7pt 
\font\fivemib=cmmib10 at 5pt  
\font\tenbsy=cmbsy10
\font\sevenbsy=cmbsy10 at 7pt 
\font\fivebsy=cmbsy10 at 5pt  
\def\BMfont{\textfont0\tenbf \scriptfont0\sevenbf
                              \scriptscriptfont0\fivebf
            \textfont1\tenmib \scriptfont1\sevenmib
                               \scriptscriptfont1\fivemib
            \textfont2\tenbsy \scriptfont2\sevenbsy
                               \scriptscriptfont2\fivebsy}
\def\BM#1{\rlx\ifmmode\mathchoice
                      {\hbox{$\BMfont#1$}}
                      {\hbox{$\BMfont#1$}}
                      {\hbox{$\scriptstyle\BMfont#1$}}
                      {\hbox{$\scriptscriptstyle\BMfont#1$}}
                 \else{$\BMfont#1$}\fi}

\def\inbar{\vrule height1.5ex width.4pt depth0pt}
\def\sinbar{\vrule height1ex width.35pt depth0pt}
\def\ssinbar{\vrule height.7ex width.3pt depth0pt}
\font\cmss=cmss10
\font\cmsss=cmss10 at 7pt
\def\ZZ{\rlx\leavevmode
             \ifmmode\mathchoice
                    {\hbox{\cmss Z\kern-.4em Z}}
                    {\hbox{\cmss Z\kern-.4em Z}}
                    {\lower.9pt\hbox{\cmsss Z\kern-.36em Z}}
                    {\lower1.2pt\hbox{\cmsss Z\kern-.36em Z}}
               \else{\cmss Z\kern-.4em Z}\fi}
\def\Ik{\rlx{\rm I\kern-.18em k}}  
\def\IC{\rlx\leavevmode
             \ifmmode\mathchoice
                    {\hbox{\kern.33em\inbar\kern-.3em{\rm C}}}
                    {\hbox{\kern.33em\inbar\kern-.3em{\rm C}}}
                    {\hbox{\kern.28em\sinbar\kern-.25em{\rm C}}}
                    {\hbox{\kern.25em\ssinbar\kern-.22em{\rm C}}}
             \else{\hbox{\kern.3em\inbar\kern-.3em{\rm C}}}\fi}
\def\IP{\rlx{\rm I\kern-.18em P}}
\def\IR{\rlx{\rm I\kern-.18em R}}
\def\IN{\rlx{\rm I\kern-.20em N}}
\def\Ione{\rlx{\rm 1\kern-2.7pt l}}
\def\IQ{\rlx\leavevmode
             \ifmmode\mathchoice
                    {\hbox{\kern.33em\inbar\kern-.3em{\rm Q}}}
                    {\hbox{\kern.33em\inbar\kern-.3em{\rm Q}}}
                    {\hbox{\kern.28em\sinbar\kern-.25em{\rm Q}}}
                    {\hbox{\kern.25em\ssinbar\kern-.22em{\rm Q}}}
             \else{\hbox{\kern.3em\inbar\kern-.3em{\rm Q}}}\fi}
\def\II{\rlx{\rm I\kern-.18em I}}

\def\scst{\scriptstyle}
\def\itrema{$\ddot{\scriptstyle 1}$}
\def\Bo{\bo{\hskip 0.03in}}
\def\lrad#1{ \left( A {\buildrel\leftrightarrow\over D}_{#1} B\right) }
\def\derx{\partial_x} \def\dery{\partial_y} \def\dert{\partial_t}
\def\Vec#1{{\overrightarrow{#1}}}
\def\.{.$\,$}

\def\grg#1#2#3{Gen.~Rel.~Grav.~{\bf{#1}} (19{#2}) {#3} }
\def\pla#1#2#3{Phys.~Lett.~{\bf A{#1}} (19{#2}) {#3}}

\def\ula{{\underline a}} \def\ulb{{\underline b}} \def\ulc{{\underline c}}
\def\uld{{\underline d}} \def\ule{{\underline e}} \def\ulf{{\underline f}}
\def\ulg{{\underline g}} \def\ulm{{\underline m}}
\def\uln#1{\underline{#1}}
\def\ulp{{\underline p}} \def\ulq{{\underline q}} \def\ulr{{\underline r}}

\def\hatm{\hat m}\def\hatn{\hat n}\def\hatr{\hat r}\def\hats{\hat s}
\def\hatt{\hat t}

\def\plpl{{+\!\!\!\!\!{\hskip 0.009in}{\raise -1.0pt\hbox{$_+$}}
{\hskip 0.0008in}}}

\def\mimi{{-\!\!\!\!\!{\hskip 0.009in}{\raise -1.0pt\hbox{$_-$}}
{\hskip 0.0008in}}}

\def\items#1{\\ \item{[#1]}}
\def\ul{\underline}
\def\un{\underline}
\def\-{{\hskip 1.5pt}\hbox{-}}

\def\kd#1#2{\d\du{#1}{#2}}
\def\fracmm#1#2{{{#1}\over{#2}}}
\def\footnotew#1{\footnote{\hsize=6.5in {#1}}}

\def\low#1{{\raise -3pt\hbox{${\hskip 1.0pt}\!_{#1}$}}}

\def\ip{{=\!\!\! \mid}}
\def\unb{{\underline {\bar n}}}
\def\upb{{\underline {\bar p}}}
\def\um{{\underline m}}
\def\up{{\underline p}}
\def\Phib{{\Bar \Phi}}
\def\Phit{{\tilde \Phi}}
\def\Phibt{{\tilde {\Bar \Phi}}}
\def\Db{{\Bar D}_{+}}
\def\gg{{\hbox{\sc g}}}
\def\nt{$~N=2$~}

\def\framing#1{\doit{#1}
{\framingfonts{#1}
\border\headpic
}}

\framing{0}

{}~~~

{\hbox to\hsize{February 1994\hfill UMDEPP 94--96}}\par

\begin{center}
\vskip  0.1in

{\large\bf Supersymmetry Breakings and}\\
\vskip 0.01in
{\large\bf Fermat's Last Theorem}$\,$\footnote{This
work is supported in part by NSF grant \# PHY-91-19746.} \\[.1in]

{\baselineskip 10pt

\vskip 0.38in

Hitoshi ~NISHINO\footnote{E-Mail: Nishino@UMDHEP.umd.edu} \\[.2in]
{\it Department of Physics} \\ [.015in]
{\it University of Maryland at College Park}\\ [.015in]
{\it College Park, MD 20742-4111, USA} \\[.1in]
and\\[.1in]
{\it Department of Physics and Astronomy} \\[.015in]
{\it Howard University} \\[.015in]
{\it Washington, D.C.~20059, USA} \\[.18in]

}
\vskip 0.8in

\end{center}
\vskip 0.3in

{\oddsidemargin=-0.2in
\evensidemargin=0.6in
\hsize=6.1in
\textwidth=6.1in

{

\noindent {\bf Abstract.}~~A mechanism of supersymmetry breaking in
two or four-dimensions is given, in which the breaking is
related to the Fermat's last theorem.  It is shown
that supersymmetry is exact at some irrational number points in parameter
space,
while it is broken at all rational number points except for the origin.
Accordingly, supersymmetry is exact {\it almost everywhere}, as well as broken
{\it almost everywhere} on the real axis in the parameter space at the same
time.  This is the first explicit mechanism of supersymmetry breaking with an
arbitrarily small change of parameters around any exact supersymmetric
model, which is possibly useful for realistically small non-perturbative
supersymmetry breakings in superstring model building.  As a byproduct, we
also give a convenient superpotential for supersymmetry breaking only
for irrational number parameters.  Our superpotential can be added as a
``hidden'' sector to other useful supersymmetric models.
}
}

\vskip 0.15in


\newpage


\def\Dot#1{\buildrel{_{_{\hskip 0.01in}\bullet}}\over{#1}}
\def\dt#1{\Dot{#1}}
\def\gg{{\hbox{\sc g}}}
\def\nt{$~N=2$~}
\def\gg{{\hbox{\sc g}}}
\def\nt{$~N=2$~}
\def\tr{{\rm tr}}
\def\Tr{{\rm Tr}}
\def\mpl#1#2#3{Mod.~Phys.~Lett.~{\bf A{#1}} (19{#2}) #3}
\def\hati{{\hat i}} \def\hatj{{\hat j}} \def\hatk{{\hat k}}
\def\hatl{{\hat l}}

\oddsidemargin=0.03in
\evensidemargin=0.01in
\hsize=6.5in
\textwidth=6.5in

\centerline{\bf 1.~~Introduction}
\vskip 0.08in

Supersymmetric theory in two-dimensions $~(D=2)$~ has interesting features
related to superconformal field theory and superstring theory, because their
phase structures can be described well by Landau-Ginzburg and Calabi-Yau
hypersurface models [1], which are easy to analyze.  It is known that the
change
of physical aspects of the theory including the topological changes of the
target space-time occurs, when the parameters in these models are varied.
If we also
need to understand the target space-time supersymmetry breaking in superstring
theory for realistic phenomenology, it is imperative to comprehend the
associated
$~N=2$~ supersymmetry breaking on the world-sheet [2].  In this connection, the
study of $~N=2$~ world-sheet supersymmetry breaking in these models must be the
first step for our ultimate goal of realistic model building.

Entirely independent of this development related to superstring
theory in physics, there has been recently some excitement in mathematics about
the possible complete proof [3] for what is called ``Fermat's last
theorem'' (FLT) [4].  This theorem dictates that there exist no integral
solutions $~l,~m,~n \in \ZZ - \{ 0 \}$~ for the algebraic equation
$$l^p + m^p = n^ p ~~, ~~~~p \in \{3,4,5,\cdots \}  ~~.
\eqno(1.1) $$
Even though there seem to be small gaps in the
recent proof [3], its validity has been widely accepted
nowadays.  (We do not address ourselves to the question of the FLT
itself, but we take its validity for granted.  This principle about
mathematical
strictness is common to other formulations in physics such as
path-integrals, or renormalizations, {\it etc.})

In this paper, we give the first application of this FLT [4] to physical
models, in particular to an $~N=(2,2)$~ supersymmetric models in $~D=2$~ or
$~N=1$~ supersymmetric ones in ~$D=4$.  We use the model in ref.~[1] for
$~D=2,\,N=(2,2)$~ supersymmetric chiral supermultiplets, and show that the
supersymmetry is broken  spontaneously for some values of the parameters
involved in the model.  In particular, we confirm the interesting and peculiar
fact that the breaking occurs at points found in any arbitrarily small
neighbourhood of each exactly supersymmetric point in the parameter space.
We also give as a by-product a superpotential that gives supersymmetry
breakings for all irrational values of parameters, while it is exact for
all rational values of parameters.


\vfil\eject

\centerline{\bf 2.~~Example of Supersymmetry Breaking by FLT}
\vskip 0.08in

Our mechanism will be useful not only for models in $~D=2$, but also for
realistic superunifications in $~D=4$.  However, for a later purpose of
generalizing the superpotential into a non-polynomial one, we temporarily stick
to $~D=2$, following the notation in ref.~[1].  We also note that
because of the parallel structure between the $~D=2,\,N=(2,2)$~
and $D=4,\,N=1$~ supersymmetries, it will be straightforward to
switch to the latter notation.

Consider a $~D=2,~N=(2,2)$~ supersymmetric system [1] with seven chiral
superfields $~\Phi_i~{\scst (i~=~1,~2,~\cdots,~7)}$, and specify the total
action as
$$\li{I \equiv & \, I_K + I_W ~~,
&(2.1) \cr
I_K \equiv & \, \sum_i \int d^2 x d ^4 \theta \, \Bar \Phi_i \Phi_i \cr
= &\, \sum_i \int d^2 x \left[ - (\partial_\m\Bar A_i)(\partial^\m A_i)
+ i \Bar\chi\low{-i} \partial_\plpl \chi\low{-i}
+ i \Bar\chi\low{+i} \partial_\mimi \chi\low{+i}
+ |F_i|^2 \right] ~~,
&(2.2) \cr
I_W \equiv &\, - \int d^2 x d\theta^+ d\theta^-
W(\Phi_i)\bigg|_{\Bar\theta^+ = \Bar\theta^- = 0} - \hbox{h.c.}
\cr
= & \, - \int d^2 x \left[ \sum_i F_i \fracmm{\partial W}{\partial\Phi_i}
+ \sum_{i,j} \fracmm{\partial^2 W}{\partial\Phi_i\Phi_j} \chi\low{-i}
\chi\low{+j} \right] \Bigg|_{\theta^\pm=\Bar\theta^\pm = 0} - \hbox{h.c.}~~.
&(2.3) \cr } $$
We are following the notation in ref.~[1] except for the names
of component fields in the $~D=4$~ style, and $~\partial_\plpl  \equiv
\partial_0
+ \partial_1,~ \partial_\mimi\equiv \partial_0 -  \partial_1$.  We specify
superpotential $~W$~ as $$ M^{-1} W (\Phi_i) = \Phi_4 \left[ \, \Phi_1^p +
\Phi_2^p - (t\Phi_3)^p \, \right] + \Phi_5 \fracmm{\sin(\pi\Phi_1)}{\pi\Phi_1}
+ \Phi_6 \fracmm{\sin(\pi \Phi_2)}{\pi\Phi_2}
+ \Phi_7 \fracmm{\sin(\pi \Phi_3)}{\pi\Phi_3} ~~,
\eqno(2.4) $$
where $~M\neq 0$~ is a real number supplying a mass dimension, and
$$t \in \IR - \{ 0 \}~~
\eqno(2.5) $$
is a parameter.  (There are many other ways of putting such
parameters.  For example, we can have $~r,~s$~
such that the first term in (2.4) is $~\Phi_4 \left[(r\Phi_1)^p +
(s\Phi_2)^p - (t\Phi_3)^p \right]$.)  Note that our superpotential
$~W(\Phi_i)$~ is regular even at $~\Phi_i=0$, due to the property of the
function $~(\sin x)/x$.  The corresponding bosonic potential $~V$~ in
component is obtained as usual by eliminating the $~F\-$auxiliary field:

\vfill\eject

$$\eqalign{M^{-2} V = \, & \sum_i \left| \fracmm{\partial W}{\partial
\Phi_i} \right|^2 \bigg|_{\theta^\pm = \Bar\theta^\pm = 0}  \cr
= \, & + \bigg| A_1^p + A_2^p - (t A_3)^p \bigg|^2 + \left| \fracmm{\sin(\pi
A_1)}{\pi A_1} \right|^2  + \left| \fracmm{\sin(\pi A_2)}{\pi A_2} \right|^2
+ \left| \fracmm{\sin(\pi A_3)}{\pi A_3} \right|^2 \cr
&+ \left| p A_4 A_1^{p-1} + A_5 \fracmm{\cos(\pi A_1)}{A_1}
- A_5 \fracmm{\sin(\pi A_1)}{\pi A_1^2} \right|^2  \cr
&+ \left| p A_4 A_2^{p-1} + A_6 \fracmm{\cos(\pi A_2)}{A_2}
- A_6 \fracmm{\sin(\pi A_2)}{\pi A_2^2} \right|^2  \cr
&+ \left| pt A_4 (t A_3)^{p-1} + A_7 \fracmm{\cos(\pi A_3)}{A_3}
- A_7 \fracmm{\sin(\pi A_3)}{\pi A_3^2} \right|^2  ~~, \cr }
\eqno(2.6) $$

We now try to minimize the potential (2.6) to see the supersymmetry for the
vacuum.  First of all we can easily see that the last six terms can vanish by
the vacuum expectation values (v.e.v.s)
$$\li{& A_1 = l~~, ~~~~ A_2 = m~~, ~~~~ A_3 = n~~;~~~~l,m,n \in \ZZ - \{ 0
\}~~,
&(2.7) \cr
& A_4 = 0 ~~, ~~~~A_5 = 0 ~~, ~~~~A_ 6= 0 ~~, ~~~~A_7 = 0 ~~.
&(2.8) \cr } $$
Now the only remaining term is
$$\left| A_1^p + A_2^p - (t A_3)^p \right|^2
\eqno(2.9) $$
If we also use the v.e.v.s (2.7) here, our question is whether or not we can
satisfy the following equation:
$$\left| l^p + m^p - (t n)^p \right|^2 = 0~~.
\eqno(2.10) $$
It is not difficult at all to solve this for ~$t$~ as
$$ t = \fracmm 1 n\left( l^p + m^p \right)^{1/p} = \fracmm l n \left[  1+
\left(\fracmm m l \right)^p \right]^{1/p} \equiv t(l,m,n) ~~.
\eqno(2.11) $$
As long as the real number ~$t$~ is chosen such that (2.11) is satisfied, the
model has a supersymmetric vacuum.  However, a simple consideration of the FLT
with (2.10) reveals that there is some more meaning in this equation, which
turns out to be exciting.

To analyze the algebraic meaning of (2.10), we develop a useful
corollary of the FLT.  We can easily prove that the FLT also implies that
the algebraic equation
$$x^p + y^p = (t z)^p~~, ~~~~ p \in \{3,4,5,\cdots \}~~,~~~~t\in\IQ-\{ 0\}~~
\eqno(2.12) $$
has {\it no} integral number solutions $~x,\,y,\,z \in \ZZ - \{ 0 \}$.   This
can be easily proved by inserting hypothetical rational number $~t =
a/b~(a,b \in \ZZ - \{ 0 \})$~ into  (2.12), and multiply both sides by $~b^p$.
The result is obviously incompatible with the FLT.

By the use of this corollary, it is now obvious that if $~t\in \IQ - \{ 0 \}$,
there are no solutions for $~A_1,\,A_2,\,A_3$, that can put (2.10)
to zero.  In other words, the vacuum of the system breaks supersymmetry for
any choice of $~t\in\IQ - \{ 0 \}$.

On the other hand, we know that supersymmetric vacuum is realized for $~t =
t(l,m,n)$~ in (2.11).  (From now on, $~t(l,m,n)$~ always denotes
(2.11), avoiding the messy expression.)  Notice the important point here that
since $~l,\,m,\,n$~ are all arbitrary non-zero integers, $~t = t(l,m,n)$~
can be made arbitrarily near to any rational number.  As a matter of
fact, we can prove this rather easily, as follows.  Let $~u \equiv
L/N \hbox{$>0$},~(L,N \in \ZZ - \{ 0 \})$~ be a positive arbitrary rational
number.  (The case of $~u<0$~ can be also proven in a similar
way.)  Then for any arbitrarily small positive real number $~\e >0$, we can
show
the existence of $~t(l,m,n)$~ such that
$$u < t(l,m,n) = \fracmm l n
\left[ 1 + \left( \fracmm m l\right)^p \right]^{1/p} < u + \e
\eqno(2.13) $$
for an appropriate choice of $~(l,m,n) \in (\ZZ - \{ 0 \})^3$.
We can first choose $~l= K L,~n = K N$~ for some large positive integer $~K$.
We next choose $~m$~ and the appropriately large enough integer $~K$~
satisfying      $$ 0 < \left(\fracmm m {K L}\right)^p \fracmm L {p N} < \e
\eqno(2.14) $$
for any small $~\e >0$.  Because this implies that
$$ 1 < 1 + \left(\fracmm m {K L}\right)^p < 1 + \fracmm p u
\e < \left(1+\fracmm\e u\right)^p = \fracmm 1 {u^p}
\left(u + \e \right)^p~~,
\eqno(2.15)$$
yielding (2.13).

Let us now introduce two sets for the parameter $~t$, depending on the
supersymmetry of the vacuum:  The set of ``supersymmetric parameters''
$$ S\equiv \left\{ t \bigg|\, t = n^{-1} \left(l^p + m^p\right)^{1/p} \in \IR
- \{ 0 \}, ~~\forall (l,m,n) \in (\ZZ - \{ 0 \})^3  \right\}~~,
\eqno(2.16) $$
and the set of ``non-supersymmetric parameters'' $~B\equiv \IR - \{ 0 \}
- S$.  Obviously $~S \cap
B = \emptyset$, and $~S \cup B = \IR - \{ 0 \}$.  Additionally, $~B
\supset \IQ - \{ 0 \}$, according to the FLT.

Remarkably, all the points in the set $~S$~ are countable and isolated, but
yet they are covering {\it almost everywhere} on the real number field,
as we have seen above.
Moreover, for any neighbourhood of an arbitrary supersymmetric model for $~t
\in S$, there exists infinitely many broken supersymmetric models for $~t'\in
B$, and {\it vice versa}!  This means that we can break any supersymmetric
model
by arbitrarily small magnitude, by choosing appropriate $~t'\in B$~ found in
any
small neighbourhood of any exactly supersymmetry parameter point
$~t\in S$.

The peculiar feature of the dependence of the ``rationality'' of
the parameter $~t$~ is rather unexpected by the general wisdom of
supersymmetry breaking based on the Witten's index $~\Tr(-1)^F$ [5].  Because
usually any small ``continuous'' change of parameters in the system, such as
from irrational numbers to rational numbers, does not trigger any supersymmetry
breaking [5].  This has been also the general principle for the renormalization
effects, where the quantum corrections will preserve the classical
supersymmetry.  For our peculiar models, supersymmetry is broken
rather ``frequently'' in the parameter space, each time the
parameters deviate from a point in $~S$~ to a point in $~B$.  This apparent
discrepancy from the usual wisdom seems to  be attributed
to the following aspects in our models.  First of all, the Witten's index is
ill-defined for our models due to the presence of a massless superfield, as
well
as the infinite degeneracy [5], as will be seen in the next section.  Second,
we expect that the index might have some implicit dependence on the
``rationality'' of the parameters.  These aspects enabled the models to escape
from the topological constraints of supersymmetry breaking, which usually
forbids such a peculiar fashion as the dependence on ``rationality'' of
parameters.  Additionally, the degeneracy of the vacua prevents us from
performing  analysis for renormalization group flows similar to that in
ref.~[6].

\bigskip\bigskip\bigskip

\centerline{\bf 3.~~Mass Spectrum around Supersymmetric Vacuum}
\vskip 0.08in

To understand our model better, we next study the mass spectrum of the system,
when there exist supersymmetric vacuum solutions.  To this end, we require
$~t\in ~S$~ in this section, satisfying (2.11) or equivalently
$$ l^p + m^p = (t n)^p~~, ~~~~p \in \{ 3,4,5,\cdots \} ~~.
\eqno(3.1) $$

As is easily seen, there can be infinitely many other
solutions for the v.e.v.s of $~A_1,~A_2,~A_3$, once there exists one set of
solutions ~$(l,\,m,\,n)$, because if we re-scale it as $~l' = q l,~m'= q m,~n'
=
q n, ~\forall q\in \ZZ - \{ 0 \}$, the new set $~(l',\,m',\,n')$~ also
satisfies (2.10).  To put it differently, we can first choose one arbitrary set
$~l,\,m,\,n \in \ZZ - \{ 0 \}$, while keeping the parameters of the
model to be the same value as the original value: $~t= t(l,m,n) =
t(l',m',n')$.
Thus this model has infinitely many supersymmetric
vacua at $~A_1 = q l,~A_2 = q m,~A_3 = q n,~ \forall q\in \ZZ
- \{ 0 \}$.  (There may be even other solutions than these, which we do
not care about so much here.)

The mass spectrum of the superpotential (2.4) around the supersymmetric
v.e.v.s $~\Phi_1 = l,~
\Phi_2 = m,~\Phi_3 = n$~ under the condition (3.1) can be easily
analyzed by appropriate field redefinitions.  We first expand each superfield
around their v.e.v.s, as
$$\eqalign{&\Phi_1 = l  + \varphi_1 ~~,
{}~~~~\Phi_2 = m + \varphi_2 ~~,~~~~ \Phi_3 = n + \varphi_3 ~~,\cr
& \Phi_4 =\varphi_4 ~~, ~~~~\Phi_5  =\varphi_5 ~~, ~~~~\Phi_6  = \varphi_6 ~~,
{}~~~~\Phi_7  = \varphi_7 ~~. \cr }
\eqno(3.2) $$
We can directly use the superfields,
because we are considering here a supersymmetric case with no v.e.v.s
for any $~F\,$-components of them.
The superfields $~\varphi_i$~ denote the fluctuations
around their v.e.v.s.  Here we rely on (2.8) with no v.e.v.s for
$~A_4, ~\cdots,~A_7$, and the stability of these solutions will be
confirmed later as the absence of tachyons.  Using also the expansion
of the function $~(\sin x)/x$, we easily get the quadratic part of $~W$:
$$ W^{(2)} = 2 \varphi_1 (a \varphi_4+ d \varphi_5)
+2 \varphi_2 (b \varphi_4 + e \varphi_6)
+2 \varphi_1 (c \varphi_4 + f \varphi_7) ~~,
\eqno(3.3) $$
where
$$\eqalign{& a \equiv \half p l^{p-1}   ~~,
{}~~~~         b \equiv \half p m^{p-1}   ~~,
{}~~~~         c \equiv \half p (t n)^{p-1}   ~~,
{}~~~~ \cr
& d \equiv \fracmm {(-1)^l}{2l} ~~, ~~~~
  e \equiv \fracmm {(-1)^m}{2m} ~~, ~~~~
  f \equiv \fracmm {(-1)^n}{2n} ~~. \cr }
\eqno(3.4) $$
All the tadpole terms linear in $~\varphi_i$~ have disappeared in $~W$~ because
of (3.1).  After the superfield redefinitions
$$\eqalign{&\varphi_1 \equiv \half \left( \Tilde\varphi_1 + \Tilde \varphi_5
 \right) ~~,  ~~~~
\varphi_5 \equiv \half d^{-1} \left( \Tilde\varphi_1 - \Tilde
 \varphi_5 - 2 a \varphi_4 \right) ~~, \cr
&\varphi_2 \equiv \half \left( \Tilde\varphi_2 + \Tilde \varphi_6
 \right) ~~,  ~~~~
\varphi_6 \equiv \half e^{-1} \left( \Tilde\varphi_2 - \Tilde
 \varphi_6 - 2 b \varphi_4 \right) ~~, \cr
&\varphi_3 \equiv \half \left( \Tilde\varphi_3 + \Tilde \varphi_7
 \right) ~~,  ~~~~
\varphi_7 \equiv \half f^{-1} \left( \Tilde\varphi_3 - \Tilde
 \varphi_7 - 2 c \varphi_4 \right) ~~, \cr }
\eqno(3.5) $$
we get
$$W^{(2)} = \half M \left( \Tilde\varphi_1^2 + \Tilde\varphi_2^2 +
\Tilde\varphi_3^2 - \Tilde\varphi_5^2 - \Tilde\varphi_6^2
- \Tilde\varphi_7^2  \right)~~,
\eqno(3.6) $$
up to some appropriate but non-essential normalization for each field.
Since (3.6) is for superpotential mass terms, the signature of each term does
{\it not} matter, and the absence of tachyons or tadpoles is also guaranteed.
The point here is that the $~\varphi_4\-$superfield has a {\it zero} mass, and
all the  other superfields are massive.  This is also consistent with our
initial  assumption
about the absence of other v.e.v.s for $~\varphi_4,\,\cdots,\,\varphi_7$.
The presence of massless chiral field $~\varphi_4$~ causes practical
difficulty when computing the Witten's index $~\Tr (-1)^F$~ [5].  The trouble
is
also caused by the enormous degeneracy related to the rescalings of
$~(l,\,m,\,n)$.

The degeneracy with respect to supersymmetric vacua can be also
seen by searching for the valleys of our potential (2.6), including also
higher order terms in addition to the mass terms above.  For example, as long
as
the conditions
$$ A_5 = (-1)^{l-1} p l^p A_4 ~~, ~~~~
A_6 = (-1)^{m-1} p m^p A_4 ~~, ~~~~A_7 = (-1)^{n-1} p (t n)^p A_4 ~~,
\eqno(3.7) $$
are satisfied, the $~A_4\-$field can be arbitrarily large, keeping the
potential
(2.6) to be zero.  In other words, there exists such a valley in the
potential, and the v.e.v.s for the $~A_4\-$field is indefinite at the classical
level.

Notice that nothing is too particular about (canonical or path-integral)
quantization around the supersymmetric vacuum.  This is because our
action (2.1) has ordinary kinetic and mass terms, but all the peculiar
effect came from higher order terms in the function $~(\sin x)/x $~ rather
``non-perturbatively''.

It is usually believed that the essential features of
$~D=2$~ supersymmetric systems, such as for relevant or marginal
operators [6], are determined by the lowest-order terms like the mass terms
or the cubic interactions, but our superpotential $~W$~ does not obey this
tendency.  In this sense, we regard the effects by the higher-order terms
as ``non-perturbative'' ones, because those infinitely higher-order terms can
not
be reached by summing up any finite number of terms.

\bigskip\bigskip\bigskip

\centerline{\bf 4.~~Other Examples of Superpotentials}
\vskip 0.08in

The mechanism proposed in this paper provides us with other
interesting by-products than the FLT itself, such as superpotentials that have
supersymmetric vacua only for rational v.e.v.s. of some bosonic field.

Take for example, a superpotential $~W\low Q$~ defined by
$$ W\low Q (\Phi, \Tilde\Phi_1, \Tilde\Phi_2, \Tilde\Phi_3) \equiv
\Tilde\Phi_2 \fracmm{\sin(\pi\Phi\Tilde\Phi_1)} {\pi\Phi\Tilde\Phi_1}
+ \Tilde \Phi_3 \fracmm{\sin(\pi\Tilde\Phi_1)}{\pi\Tilde\Phi_1} ~~.
\eqno(4.1) $$
The corresponding bosonic potential is
$$\eqalign{V\low Q = \, & + \left| \fracmm{\sin(\pi A \Tilde A_1)}
{\pi A \Tilde A_1} \right|^2 + \left| \fracmm{\sin(\pi\Tilde A_1)}
{\pi\Tilde A_1} \right|^2
+ \left|\Tilde A_2 \fracmm{\sin (\pi A \Tilde A_1) - \pi A \Tilde A_1
\cos(\pi A\Tilde A_1)}{\pi A^2\Tilde A_1} \right|^2  \cr
& + \left| \Tilde A_2 \fracmm{\sin (\pi A \Tilde A_1) - \pi A \Tilde A_1
\cos(\pi
A\Tilde A_1 )}{\pi A\Tilde A_1^2} + \Tilde A_3 \fracmm{\sin (\pi\Tilde A_1) -
\pi\Tilde A_1 \cos(\pi\Tilde A_1 )} {\pi\Tilde A_1^2}  \right|^2  ~~. \cr}
\eqno(4.2) $$

The last two terms in (4.2) can be made zero by the v.e.v.s
$$ \Tilde A_2 = \Tilde A_3 = 0 ~~,
\eqno(4.3) $$
while the first two terms will vanish only at
$$\Tilde A_1 \equiv n~~, ~~~~ A \Tilde A_1 \equiv l ~~,~~~~(n,~l \in
\ZZ -\{ 0 \})~~.
\eqno(4.4) $$
This implies that
$$ A = \fracmm l n \in \IQ - \{ 0 \} ~~ ,~~~~A_1\in \ZZ - \{ 0 \}~~,
\eqno(4.5) $$
in particular, the v.e.v.s~of $~A$~ must always be a non-zero {\it rational
numbers}.

We have thus seen that the superpotential $~W\low Q$~ is generating rational
number v.e.v.~s for the $~A\-$component of the chiral superfield $~\Phi$.  The
other three {\it tilded} superfields are playing the role of auxiliary
superfields, and $~V\low Q$~ can be minimized at $~\Tilde A_1 = \Tilde A_2=
\Tilde A_3 = 0$.

As for the Witten's index of this model, it is ill-defined due to the
``valley''
structure of the potential $V\low Q$~ along the direction of an arbitrary large
value of $~A_3$~ as seen from the last term in (4.2).  Thus we can not see
the conservation of topology depending on the value of the v.e.v.s.

An interesting application of $~W\low Q$~ is the following superpotential:
$$ W_1 = \Phi_3 (\Phi_1 - r) + \Phi_4 (\Phi_2 - s) + W\low Q
(\Phi_1,\Tilde\Phi_5,\Tilde\Phi_6,\Tilde\Phi_7)
+ W\low Q (\Phi_2,\Tilde\Phi_8,\Tilde\Phi_9,\Tilde\Phi_{10})~~,
\eqno(4.6) $$
yielding the bosonic potential
$$\eqalign{ V_1 \equiv \, & + \left| A_1 - r \right|^2 + \left| A_2
- s \right|^2
+\left| f(\pi\Tilde A_5) \right|^2 + \left| f(\pi\Tilde A_8) \right|^2 \cr
& + \left|\Tilde A_6 g(\pi A_1 \Tilde A_5) + \Tilde A_7 g(\pi \Tilde A_5)
\right|^2
  + \left|\Tilde A_9 g(\pi A_2 \Tilde A_8) + \Tilde A_7 g(\pi\Tilde A_8)
\right|^2  \cr
& + \left| \Tilde A_6 g(\pi A_1 \Tilde A_5) + A_3  \right|^2 + \left| \Tilde
A_9 g(\pi A_2 \Tilde A_9) + A_4  \right|^2   ~~, \cr}
\eqno(4.7) $$
where
$$ f(x) \equiv \fracmm {\sin x} x ~~, ~~~~ g(x) \equiv f'(x) \equiv
\fracmm{x \cos x - \sin x}{x^2}~~.
\eqno(4.8) $$

As before, the 2nd.~and 3rd.~lines in (4.7) vanish at the v.e.v.s
$$ A_3 = A_4 = \Tilde A_6 = \Tilde A_7 = \Tilde A_9 = \Tilde A_{10} = 0~~,
\eqno(4.9) $$
while the 3rd.~and 4th.~term vanish, when
$$ A_1 \Tilde A_5 \equiv m~~, ~~~~ A_2 \Tilde A_8 \equiv n ~~,
{}~~~~\Tilde A_5 \equiv k ~~, ~~~~ \Tilde A_8 \equiv l~~, ~~~~
(k,l,m,n \in\ZZ - \{ 0 \}) ~~,
\eqno(4.10) $$
or equivalently,
$$\eqalign{ & A_1\, , ~ A_2 \in \IQ - \{ 0 \} ~~, \cr
&\Tilde A_5\, , ~ \Tilde A_8  \in \ZZ - \{ 0 \} ~~. \cr }
\eqno(4.11) $$

Considering these with the remaining first two terms in (4.7) easily reveals
that

\Item{(i)} $~\forall (r,s) \in (\IQ - \{ 0 \} )^2 \equiv S \Longrightarrow
V_{\rm min} = 0$:~~supersymmetric vacuum

\Item{(ii)} $~\forall (r,s) \in \IR^2 - S \equiv B
\Longrightarrow V_{\rm min} >
0$:~~non-supersymmetric vacuum

\noindent The interesting feature here is that depending on the ``rationality''
of the parameters $~(r,s)$, the system has either supersymmetric or
non-supersymmetric vacua.  Needless to say, we could have chosen only $~r$~ as
a
one-dimensional parameter, or as many as we wish like $~(r_1,r_2,\cdots,r_n)$~
by adding $~n$~ copies of $~W\low Q$.

\bigskip\bigskip\bigskip

\centerline {\bf 5.~~Concluding Remarks}
\vskip 0.08in

In this paper we have presented an explicit mechanism characterized by the
superpotential (2.4), in which supersymmetry breakings occur
with {\it arbitrarily small} changes of parameters around isolated exact
supersymmetric models, depending on the ``rationality'' of the parameter $~t$.
On the real axis in the parameter space of $~t$, supersymmetry is found to be
exact {\it almost everywhere}, as well as is spontaneously broken
{\it almost everywhere}, at the same time.  We believe the validity of our
result, as long as the FLT [3,4] is acceptable.

At first sight, there appeared to be an incompatibility of this result with
the general wisdom about the non-zero Witten's index [5] of a supersymmetric
model.  We understand that this is attributed to the ill-defined
Witten's index of
our model due to the massless superfield, and also to its possible dependence
on
the ``rationality'' of the parameter $~t$~ in some implicit way.

One of the interesting aspects of our model is the possibility
of arbitrarily small supersymmetry breaking.  This is because the
breaking scale can be made as small as we wish, due to the ``arbitrarily''
small
breaking  effect on the bosonic potential by shifting the parameters from
the exact supersymmetric values $~t = t(l,m,n),~\forall (l,\,m,\,n) \in
(\ZZ - \{ 0 \})^3$~ to an arbitrarily close rational numbers $~t'\in
\IQ-\{ 0 \}$~ for ~$W$.

As is well-known, there is non-renormalization theorem applied to the
$~F\-$type superpotential terms.  Considering this theorem, we can conclude
that
the peculiar structure of our superpotential will be maintained even at the
quantum level, if we have started with an exactly supersymmetric classical
vacuum.  This is the case only when the classical vacuum preserves
supersymmetry, because the non-renormalization theorem applies only to the
system with tree-level supersymmetry.  It is therefore interesting to see
what kind of quantum corrections will be generated, when the starting classical
vacuum is non-supersymmetric such as the case $~t \in \IQ - \{ 0 \}$.
It is also amusing that the supersymmetry is protecting a supersymmetric model
against any quantum perturbations, that might shift the parameters away from
the
original set $~S$~ to ``next'' infinitesimally close irrational numbers in
$~B$,
with such ``infinite'' accuracy.  Furthermore, it is especially in
supersymmetric models in which the positivity of the potential plays an
important role, because of supersymmetry breakings related to the non-zero
vacuum
energy.  In ordinary models in physics, the ``rationality'' of constants and/or
fields does not matter {\it unlike} our model, in which supersymmetry {\it
distinguishes} them.  From these viewpoints, together with the
renormalizability
for the $~D=2$~ case, we believe that our models are not just of ``accidental''
interest, but they signal more fundamentally significant connection between
supersymmetric field theories in $~D=2$~ and the FLT in number theory.

We also mention the most important practical application of our model.  Our
superpotential (2.4) can be treated as a ``hidden'' sector added to
other useful $~D=2,\, N=(2,2)$~ supersymmetric models, in order to break
supersymmetry with small magnitude.  This is because the presence of our
superfields will not interfere with the fundamental structure of other
sectors, such as the mass spectrum or manifold structures,
except for the supersymmetry breaking at a global minimum.  We are sure that
there can be more to be done for interesting applications of our models
combined
with other useful models.  Another interesting application is the $~D=4$~
locally supersymmetric unifications [7], in which the renormalizability of the
superpotential is no longer crucial, once supergravity is included.  The
usage of our superpotential as a ``hidden'' sector may well have some advantage
over the conventional Polony-type superpotential [7], due to the possibility
of small supersymmetry breaking of $~{\cal O}(10^{-15})$~ needed for realistic
model building which keeps the zero-ness of the cosmological constant.

To our knowledge, our models have provided the mechanism which presents a
peculiar link between the FLT in number theory [3,4] and the vacuum structure
of
supersymmetry in such an explicit way for the first time.  The only well-known
connection between number theory and supersymmetry has been {\it via}
topological effects, such as instantons and monopoles in supersymmetric
models.  (However, see ref.~[8] in which string coupling constant is
parametrized by rational numbers.)  Traditionally, supersymmetry has been
always supposed to act on
general real (or complex) fields rather continuously without distinguishing
rational number parameters from irrational ones.  We believe that our models
have opened a new direction to the studies of such an important issue as
supersymmetry breaking for the purpose of realistic model building as well as
for purely mathematical or theoretical interest.

We are indebted to W.W.~Adams, S.J.~Gates, Jr., T.~H\"ubsch, and J.~Swank for
valuable discussions.

\vfill\eject

\refs
\small

\def\\{\vskip 0.05in}
\def\item#1{\Item{#1}}
\def\items#1{\\ \item{{#1}.}}

\items{1} E.~Witten, \np{403}{93}{159}.

\items{2} See e.g., M.~Green, J.H.~Schwarz and E.~Witten, {\it Superstring
Theory}, Vols.~I and II, Cambridge University Press (1987).

\items{3} Andrew Wiles, to be published.

\items{4} P.~de Fermat,
{\it `Observatio'} in ``{\it Arithmetica of Diophantus}'' (1621),
unpublished (1637);
\item{  } K.A.~Ribet, Notices of Amer.~Math.~Soc.~{\bf 40} (1993) 575;
\item{  } D.A.~Cox, Amer.~Math.~Monthly {\bf 101} (1994) 3.

\items{5} E.~Witten, \np{202}{82}{253}.

\items{6} C.~Vafa and N.P.~Warner, \pl{218}{89}{51}.

\items{7} See, e.g., H.P.~Nilles, \prep{110}{84}{1}.

\items{8} H.~Nishino, \mpl{7}{92}{1805};
\item{  } H.~Nishino and S.J.~Gates, \ijmp{8}{93}{3371}.

\end{document}